\documentclass[pre,twocolumn,preprintnumbers,amsmath,amssymb,longbibliography]{revtex4-1}
\usepackage{amssymb,amsmath,amssymb,amsthm,graphicx}
\usepackage{amsmath,amssymb}
\usepackage{float}
\usepackage{graphicx}
\usepackage{psfrag}
\usepackage{color}
\usepackage{dcolumn}
\usepackage{bm}
\usepackage[normalem]{ulem}
\usepackage[absolute]{textpos}

\def\beq{\begin{equation}}
\def\eeq{\end{equation}}
\def\bea{\begin{eqnarray}}
\def\eea{\end{eqnarray}}
\newcommand{\stkout}[1]{\ifmmode\text{\sout{\ensuremath{#1}}}\else\sout{#1}\fi}

\begin{document}
 
%\title{Hydrodynamic theory for active symmetric membranes}
\title{%Active effects can stabilise or destabilise symmetric membranes\\
Flat or crumpled: states of active symmetric membranes}
\author{Sudip Mukherjee}\email{sudip.bat@gmail.com,aca.sudip@gmail.com}
\affiliation{Barasat Government College,
10, KNC Road, Gupta Colony, Barasat, Kolkata 700124,
West Bengal, India}
\affiliation{Max-Planck-Institut f\"ur Physik komplexer Systeme, N\"othnitzer Strasse 38, 01187 Dresden,Germany}
%\affiliation{Theory Division, Saha Institute of
%Nuclear Physics, 1/AF Bidhannagar, Calcutta 700064, West Bengal, India}
\author{Abhik Basu}\email{abhik.123@gmail.com,abhik.basu@saha.ac.in}
\affiliation{Theory Division, Saha Institute of
Nuclear Physics, 1/AF Bidhannagar, Calcutta 700064, West Bengal, India}
\affiliation{Max-Planck-Institut f\"ur Physik komplexer Systeme, N\"othnitzer Strasse 38, 01187 Dresden,Germany}

\begin{abstract}

%[IT WILL BE A SHORT AND SIMPLE PAPER]

 We set up and study the hydrodynamic theory for { {inversion-symmetric}} active fluid and tethered membranes.   For some choices of the activity parameter, such membranes are stable  and described by  linear hydrodynamic equations, which are exact in the asymptotic long wavelength limit, giving stable flat phases  with translational quasi long range orders. { {For other choices of the activity parameter, the system is linearly unstable in the long wavelength limit, implying crumpling, or has intermediate wavevector instabilities, suggesting patterns.}} We argue that in such an active membrane thermal noises dominate over any active noises, and use those to calculate the correlation functions of membrane conformation fluctuations  in the stable case, and the associated correlation functions of the embedding bulk flow velocities.
\end{abstract}

\maketitle

%\section{Introduction}

%\section{Equilibrium membranes}
%\sout{Membranes form essential components of biological cells, enclosing them in finite volumes~\cite{alberts}. }
% in a tethered membrane, that take it out of the scope of MWHT.%\sout{ In addition to the their thermodynamic properties of their low-$T$ ``flat'' phase, dynamics of equilibrium tethered membrane can be used to explain the flicker phenomenon observed in the red blood cells in mammals~\cite{frey-flicker}.}

%\sout{Lipid membranes enclosing live biological cells are necessarily out of equilibrium due to the various microscopic active processes undergoing in live cells, e.g., e.g. nonequilibrium ﬂuctuations of cell
%cytoskeletons~\cite{active-cyto} and active proteins in the lipid membrane~\cite{active-protein}.  As a consequence of these active processes, }

{ {Considerable success has been achieved by now in applying the principles of thermodynamics and statistical mechanics to membranes in thermal equilibrium  in terms of continuum theories, parametrized by a set of elastic constants~\cite{nelson,udo,safran}. Such theories have successfully predicted and explained a variety of results, e.g., thermal crumpling of large tensionless lipid membranes at any low but non-zero temperature $T$~\cite{nelson,udo,luca,tirtha-pre} and the existence of statistically flat phases orientational long range order (LRO) at sufficiently low but finite $T$  in tethered membranes~\cite{nelson,kantor,lubensky}. It is however now understood that due to the underlying microscopic active processes in live cells~\cite{alberts}, e.g., nonequilibrium ﬂuctuations of cell
cytoskeletons~\cite{active-cyto} or active proteins in lipid membranes~\cite{active-protein,active-protein1}, live cell membranes display unusual dynamical fluctuation properties~\cite{fluc1,fluc2,fluc3,fluc4}. Similarly, experimental comparisons of fluctuations in healthy, live red blood cell membranes and those in ATP (adenosine triphosphate) depleted red blood cells suggest that active processes should be responsible in live red blood cell membrane fluctuations~\cite{rbc-exp1,rbc-exp2,rbc-exp3}. More recently, by comparing membrane response and fluctuations on single red blood cells, a violation of the Fluctuation–Dissipation-Theorem (FDT)~\cite{chaikin} in the flickering of live red blood cell demonstrating the underlying nonequilibrium, active processes is reported in Ref.~\cite{joanny-natphys}. The generality of experimental observations on the manifestations of nonequilibrium nature of cell membrane dynamics has prompted scientists to look for minimal physical descriptions. Such efforts have led to theories of active membranes applicable to various settings~\cite{joanny-natphys,prost,pump,micropipet,nir,nir1,nir2,udo1,sriram,tirtha-njp,tur1,liang,kulkarni}. We here construct and study the hydrodynamic equations for open, inversion-symmetric, isotropic single-component active fluid or tethered membranes, which are tensionless in the equilibrium limit and embedded in three-dimensional (3D) isotopic fluids.}} %Our results complement the existing theories. 

Our principal results are as follows: (i) The nonequilibrium dynamics of nearly flat symmetric active membranes, fluid or tethered, emerge from  their couplings with an embedding active isotropic bath, provided by a 3D active but isotropic fluid. (ii) The linearized hydrodynamic equations predict that such a nearly flat patch can be stable, corresponding to orientational LRO for a fluid membrane, or orientational LRO and in-plane translational quasi long range order (QLRO). A {\em positive active tension}, generated by the interplay between active processes and 3D embedding fluid flows, is responsible for orientational LRO. These results are {\em exact} in the asymptotic long wave length limit, with all nonlinear effects being irrelevant. Thus nonlinear effects play {\em no} role in sustaining the order. (iii) Although these results are reminiscent of those for  equilibrium fluid or tethered membranes {\em with a finite tension} at some effective temperature, in our linear theory the equal-time undulation mode correlation function depends on a dynamic (active, see below) coefficient, indicating the inherent nonequilibrium nature of the system. In contrast, the in-plane displacement correlators depend only on thermodynamic or equilibrium parameters. (iv) The undulation modes can be linearly unstable at intermediate scales { {presumably giving rise to steady patterns}}, if the membrane is permeable and the associated permeation flow is { {sufficiently strongly}} destabilizing. (v) The active tension can also be negative, in which case the undulation modes are linearly unstable at the longest scales, signalling  membrane crumpling. Interestingly, the in-plane displacement modes are still linearly stable, but should get {\em nonlinearly} unstable, possibly due to their nonlinear couplings with the undulation modes, which are irrelevant in the linearly stable case. Our theoretical predictions are general and can be tested in artificial {\em in-vitro} settings.

\begin{figure}[htb]
\includegraphics[width=7.3cm]{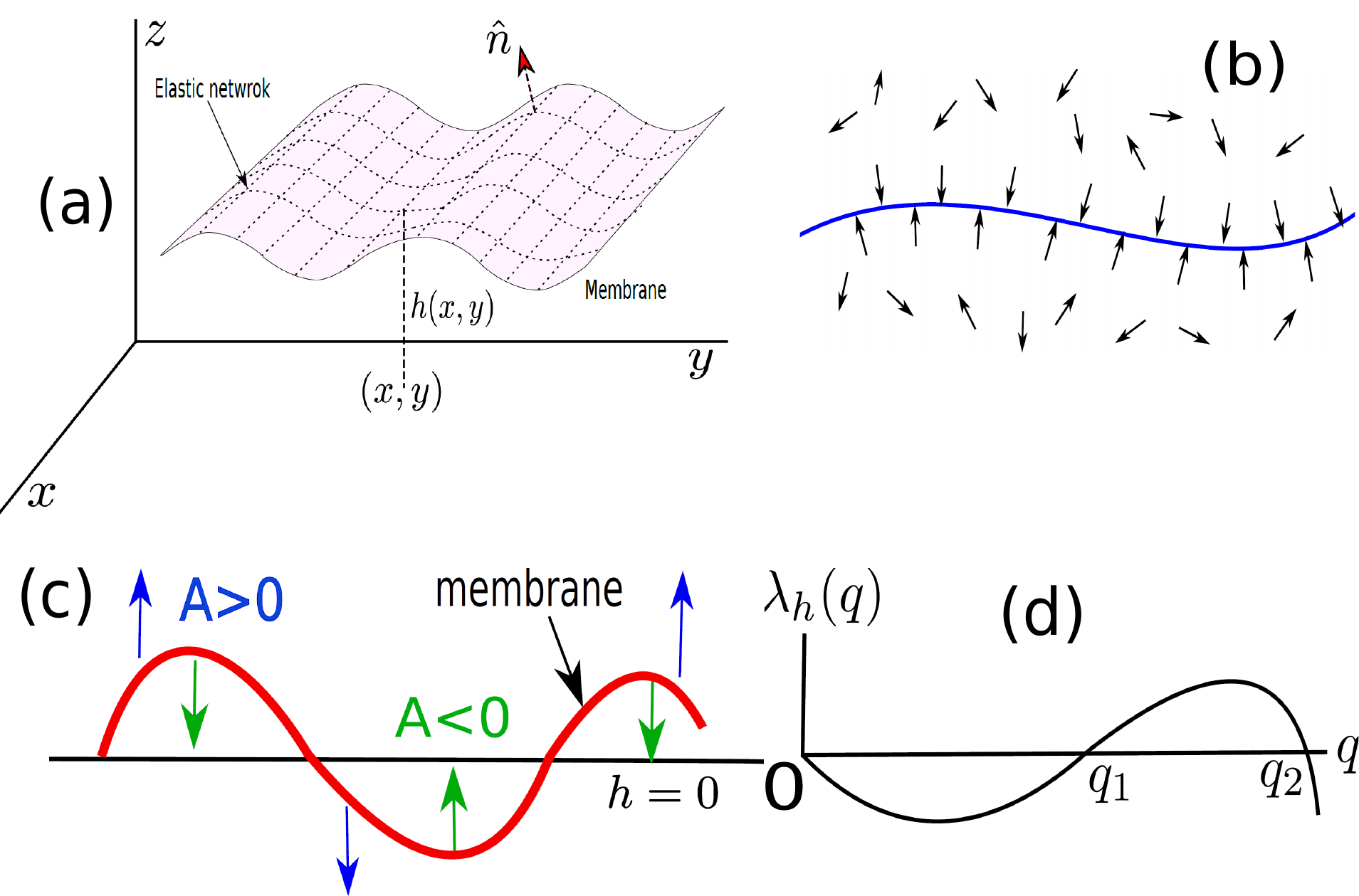}
 \caption{ Schematic (a) model diagram of a membrane depicted in the Monge gauge that is embedded in an active isotropic fluid. Here, ${\bf R}=({\bf r} + {\bf u},h({\bf r},t))$, where $h({\bf r},t)$ is the height above points ${\bf r}=(x,y)$ on a reference plane. Further, $\hat n$ is the local normal to the membrane. (b) Schematic diagram showing actin filaments grafted normally to the membrane; in the bulk they are in the isotropic phase. (c) Diagram depicting the role of active stress in stabilizing ($ {A<0}$, green arrows indicating {\em reduction} in fluctuations)/destabilizing ($ {A>0}$, blue arrows indicating {\em enhancement} of fluctuations) the membrane.  (d) plot of $\lambda_h({\bf q})$ versus $q$ showing an unstable intermediate band of wavevectors. Within the range $q_1<q<q_2$, $\lambda_h ({\bf q})>0$, implying linear instability. See text.}\label{model-fig}
\end{figure}

We now construct the hydrodynamic equations for a permeable active symmetric tethered membrane embedded in a 3D bulk isotropic fluid. { {The hydrodynamic equation for an active symmetric fluid membrane can be obtained by   ignoring the in-plane displacement modes, since these are {\em not} broken symmetry hydrodynamic modes for fluid membranes}}. The conformation ${\bf R} ({\bf s })$ of any membrane in 3D is parametrized by a two-dimensional (2D) position vector ${\bf s}$ that designates points on the membrane~\cite{convec1,convec2}.  The local membrane velocity is then formally given by $ \partial_t {\bf R}({\bf s})$. For a nearly flat patch of a membrane without any overhangs, it is convenient to use the Monge gauge~\cite{nelson,chaikin}, in which ${\bf R}=({\bf r}+{\bf u},h({\bf r},t))$, where $h({\bf r},t)$ is the height above points ${\bf r}=(x,y)$ on a reference plane; see Fig.~\ref{model-fig}(a) for a schematic diagram of the membrane. Here, ${\bf u}$ is the relative in-plane displacement of a point whose location in the undistorted membrane is at ${\bf R}=({\bf r},h=0)$. This reference plane is however completely arbitrary.

%\begin{widetext}

%\end{widetext}

We denote the 3D bulk hydrodynamic flow velocity by $v^\prime_\alpha,\,(\alpha=x,y,z)$ and the 2D flow velocity in the membrane by $v_i,\,i=x,y$. Ignoring any number conservation, $h$ and ${\bf u}\equiv u_i$ are the two hydrodynamic variables in the problem along with the total momentum of the system. In the presence of a permeation flow velocity $v_\text{perm}$ locally normal to the membrane, considered as a permeable fluid film~\cite{cai}, there is a local normal relative velocity ($=v_\text{perm}$) between the membrane and the bulk fluid flow at the location of the membrane normally to it. With $\partial_t h\equiv \partial_t R_z$ in the Monge gauge,  $h$ obeys
\begin{equation}
 \frac{\partial h}{\partial t}-v_\text{perm}=v_z^\prime|_{z=h}.\label{eq-h}
\end{equation}
In general, $v_\text{perm}$ is a function of $h$ and $u_{ii}$ that odd in $h$, tilt-invariant due to the rotational symmetry~\cite{nelson} and also invariant under $u_i\rightarrow u_i+const.$.  Since the permeation flow is locally normal to the membrane,  $v_\text{perm}=\sqrt{1+({\boldsymbol\nabla}h)^2}f_\text{perm}(\nabla^2 h,u_{ij})$, where $f_\text{perm}(\nabla^2 h,u_{ii})$ is a general function of its arguments and their in-plane space derivatives, and odd in $h$. For small fluctuations, expanding in $h$ and $u_{ii}$
\begin{equation}
 v_\text{perm}= \mu_1\nabla^2 h + \frac{\mu_1}{2} \nabla^2 h ({\boldsymbol\nabla}h)^2 + \mu_2 {\boldsymbol\nabla}\cdot {\bf u}\nabla^2 h + \mu_p \frac{\delta {\cal F}}{\delta h},\label{v-perm-eq} 
\end{equation}
retaining only the lowest order nonlinearities. Here, $u_{ij}=\frac{1}{2}(\partial_i u_j +\partial_j u_i +\partial_i h \,\partial_j h)$ %\label{strain}
%\end{equation}
is the strain tensor~\cite{nelson,chaikin},
 $\mu_p>0$ is an equilibrium permeation coefficient, $\mu_1,\,\mu_2$ give strengths of active permeation, each of which can be positive or negative. { {Indeed, symmetry
arguments, which is used to write down (\ref{v-perm-eq}), say nothing about the active parameters, whose magnitudes and signs are determined by
the microscopic physics of a particular situation.}} For instance, $\mu_1$, effectively a nonhydrodynamic tension, can be made negative by strong curvature-dependent active fusion processes on the membrane~\cite{sarosij,patricia}, or by curvature-inducing molecules~\cite{curv-ind-mols}. It can also be positive for certain protein pumps in membranes; see~Refs.~\cite{pump,noguchi}.   Here, $\cal F$ is the Landau-Ginzburg free energy functional of a tensionless tethered membrane. General symmetry considerations dictate
\begin{equation}
 {\cal F}=\frac{1}{2}\int d^2r\,[K(\nabla^2 h)^2 + \lambda u_{ii}^2 + 2\mu u_{ij}u_{ij}];\label{free-en}
\end{equation}
%\begin{equation}
with
 $K>0$  the bend modulus and $\lambda,\mu$ are the Lam\'e coefficients~\cite{chaikin}. The equation of $u_i$ is
\begin{equation}
 \frac{\partial u_i}{\partial t} = v_i.\label{basic-u-eq}
\end{equation}
In the extreme Stokesian limit, dropping the fluid inertia, 3D hydrodynamic velocity $v_\alpha^\prime$ is obtained from the force balance condition $\nabla_\beta\sigma^\prime_{\alpha\beta}=0$, where stress tensor %$\sigma_{\alpha\beta}^\prime$ %is given by
\begin{equation}
 \sigma^\prime_{\alpha\beta}=\eta'(\partial_\alpha v'_\beta+\partial_\beta v'_\alpha)-P^\prime\delta_{\alpha\beta}
\end{equation}
for an incompressible bulk fluid~\cite{landau-fluid},  $\eta'$ is the shear viscosity and $P'$ is the 3D pressure. 

Ignoring inertia, the 2D in-plane velocity $v_i$ satisfies a force balance equation. There are internal forces in the membrane to be balanced by the external forces on the membrane from the embedding 3D bulk fluid. The internal forces per unit area of the membrane are $\nabla_j\sigma^\text{tot}_{ij}$, where $\sigma^\text{tot}_{ij}$ is the total 2D internal stress:
\begin{equation}
 \sigma^\text{tot}_{ij}=\eta (\partial_i v_j+\partial_j v_i) + \eta_b {\boldsymbol\nabla}\cdot {\bf v}\delta_{ij}- P\delta_{ij} +\sigma_{ij}^\text{el}+ \sigma_{ij}^a,\label{2d-stress}
 \end{equation}
 $\eta$ and $\eta_b$ are the 2D shear and bulk viscocities respectively and $P$ is a 2D partial pressure. Further, $\sigma_{ij}^\text{el}=\delta {\cal F}/\delta u_{ij}$ is the elastic stress~\cite{chaikin,ising-elastic-long}. The last contribution to $\sigma^\text{tot}_{ij}$ comes from the active stress $\sigma_{\alpha\beta}^a=A n_\alpha n_\beta$, where $\hat {\bf n} = (\hat z,-\nabla_i h)$ is the local normal to the leading order in $h$-fluctuations with $\hat z$ being the unit vector along the $z$-direction. This active stress can locally generate forces  $\nabla_j\sigma^a_{ij}$ tangentially and $\nabla_j\sigma^a_{zj}$ normally to the membrane, and 
 can arise, e.g., in a membrane with actin filaments grafted normally to it (or also in a isotopic live bacterial suspension). The actin polymerzation or depolymerization can force the membrane~\cite{actin-force1,actin-force2}. Now define a 3D polarization vector $ {\bf p}\equiv p_\alpha$ that is non-vanishing only at the location of the membrane. Elsewhere in the bulk active fluid being in its isotropic phase, we set $ {\bf p}=0$. For a nearly flat membrane, the condition of normal anchoring gives $p_\alpha = - \hat n_\alpha$~\cite{niladri-epje3}. A non-zero $p_\alpha$ on the membrane allows us to define an active stress $\sim A p_\alpha p_\beta$~\cite{frank1,frank2,frank-phys-rep,joanny-prost,sriram-ann-rev,sriram-RMP}, which with normal anchoring gives $\sigma^a_{\alpha\beta}$ as defined above, with $A<(>)0$ corresponding to contractile, e.g., as in actin cytoskeletons (extensile, e.g., in some bacterial suspensions)~\cite{sriram-RMP} active stress. { {See Fig.~\ref{model-fig}(b) for a schematic representation.}}

The 2D force balance equation for the in-plane flow on the membrane reads~\cite{brochard,kramer,niladri-epje4}
\begin{equation}
 \nabla_j\sigma^\text{tot}_{ij} + f_i=0.\label{2d-stokes}
\end{equation}
Here, $f_i$ is the force per unit area on the 2D membrane due to the shear stress of the bulk flow:
\begin{equation}
 f_i = \eta'\big(\frac{\partial v_i'}{\partial z}+\frac{\partial v_z'}{\partial x_i}\big)|_{z=h_+}-\eta'\big(\frac{\partial v_i'}{\partial z}+\frac{\partial v_z'}{\partial x_i}\big)|_{z=h_-}.\label{shear-stress}
\end{equation}
Normal stress balance at the membrane yields
\begin{equation}
 (2\eta' \frac{\partial v_z'}{\partial z}-P')|_{z=h_+}-( 2\eta'\frac{\partial v_z'}{\partial z}-P')|_{z=h_-}-\frac{\delta {\cal F}}{\delta h} + \nabla_j \sigma^a_{zj}=0\label{normal-stress}
\end{equation}
 balancing the external forces per unit area or the normal stresses by the bulk fluid on the membrane with the internal forces per unit area by the membrane that include elastic restoring forces of the membrane and active forces due to the active stress defined above.

We impose $v_\alpha^\prime (z=\pm \infty)=0$ as the boundary conditions on $v_\alpha^\prime$. Lastly, the no-slip boundary condition on the membrane implies $v'_{\alpha=i}(z=h_+)=v'_{\alpha=i}(z=h_-)=v_i$ and the continuity condition $v^\prime_z(z=h_+)=v_z^\prime(z=h_-)$. We now solve for $v^\prime_\alpha({\bf r},z,t)$ subject to the above conditions. Defining $v_\alpha^\prime ({\bf q},z,t)$ and $v_i({\bf q},t)$ as the in-plane Fourier transforms  of $v_\alpha({\bf r},z,t)$ and $v_i({\bf r},t)$, where $\bf q$ is a wavevector, (see, e.g., Refs.~\cite{brochard,kramer,niladri-epje4}) and
%\begin{equation}
% \frac{\partial h ({\bf q},t)}{\partial t}= \frac{-A q}{4\eta'}h({\bf q},t) + v_\text{perm} - \frac{1}{4\eta' q}\frac{\delta {\cal F}}{\delta h}.\label{h-eq-gen}
%\end{equation}
by using (\ref{v-perm-eq}) (\ref{free-en}), and ignoring nonlinearities, we obtain
\begin{equation}
 \frac{\partial h ({\bf q},t)}{\partial t}= \frac{{ A} q}{4\eta'}h({\bf q},t) - \mu_1 q^2 h({\bf q},t) - \frac{Kq^3}{4\eta'} h({\bf q},t),\label{h-eq-lin}
\end{equation}
which is the linearized hydrodynamic equation for the undulation modes.%, where the term with $\mu_p>0$ is subleading to the term with $K>0$ and hence is ignored. 
The intermediate calculational details, similar to Refs.~\cite{kramer,sriram,niladri-epje4}, are straight forward, and are given in Supplemental Material (SM)~\cite{sm}.

To obtain the hydrodynamic equations for $\bf u$, we decompose ${\bf u}=(u_\parallel,\,u_\perp)$ and ${\bf v}=(v_\parallel,\,v_\perp)$, where $\parallel$ and $\perp$ refer to in-plane directions parallel and perpendicular to $\hat {\bf q}$. %\sout{With these decomposition and using (\ref{basic-u-eq}), we have}
%\begin{equation}
% \frac{\partial u_a ({\bf q},t)}{\partial t}=v_a({\bf q},t).
%\end{equation}
To calculate $v_\perp$ and $v_\parallel$, we resolve (\ref{2d-stokes})  perpendicular and parallel to $\hat {\bf q}$ giving $v_\perp ({\bf q},t)=-[\mu q/(2\eta')]\, u_\perp ({\bf q},t)$ to the linear order in (small) fluctuations. Then
\begin{equation}
 \frac{\partial u_\perp({\bf q},t)}{\partial t}= -\frac{\mu q}{2\eta'}\, u_\perp ({\bf q},t)\label{uperp-eq-lin}
\end{equation}
is the linearized hydrodynamic equation for $u_\perp$. In the long wavelength limit, the equation for $v_\parallel({\bf q},t)$ reads
\begin{equation}
 4\eta'qv_\parallel ({\bf q},t)=-iqP(\rho) - (\lambda + 2\mu)q^2 u_\parallel ({\bf q},t).\label{v-parallel-inter}
\end{equation}
Here,  $\rho$ is the 2D membrane density.  Now assume the total particle number $\int d^2 r\,\rho$ is constant only on average due to possible non-number conserving events, e.g., birth-death of the particles (assumed active). Then $\rho$ is not a conserved density and hence is not a hydrodynamic variable. In fact, fluctuations $\delta\rho$ about the mean concentration $\rho_0$ relaxes fast. In the spirit of coarse-grained hydrodynamic approaches, we then replace $\chi(\rho)$ by $\chi(\rho_0)$, a constant. This ensures that $P$ drops out of (\ref{v-parallel-inter}).  Solving for $v_\parallel({\bf q},t)$ we get
\begin{equation}
 \frac{\partial u_\parallel ({\bf q},t)}{\partial t}= -\frac{\lambda +2\mu}{4\eta'}q u_\parallel({\bf q},t)\label{u-parallel-linear}
\end{equation}
to the linear order in (small) fluctuations.

\begin{figure}[htb]
\includegraphics[width=4.1cm,height=2.5cm]{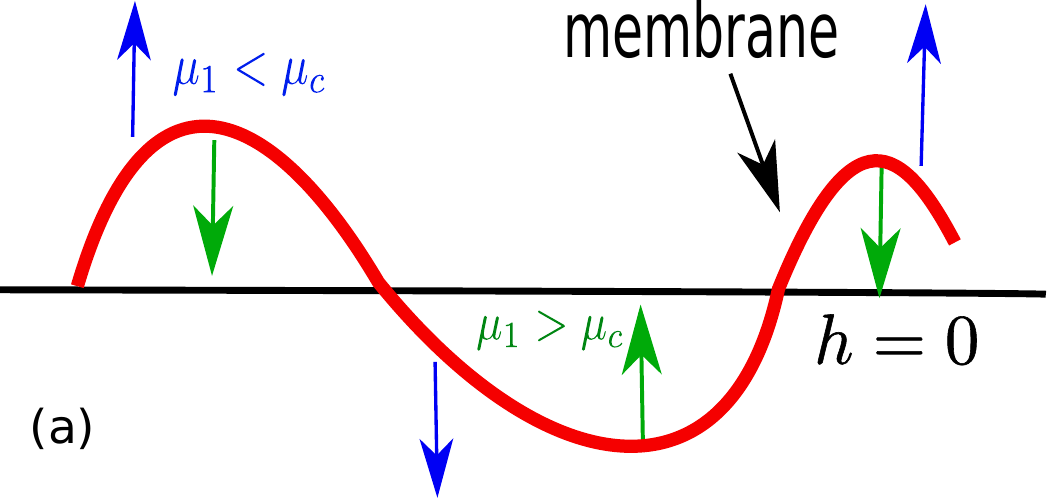}\hfill
\includegraphics[width=4.1cm,height=2.5cm]{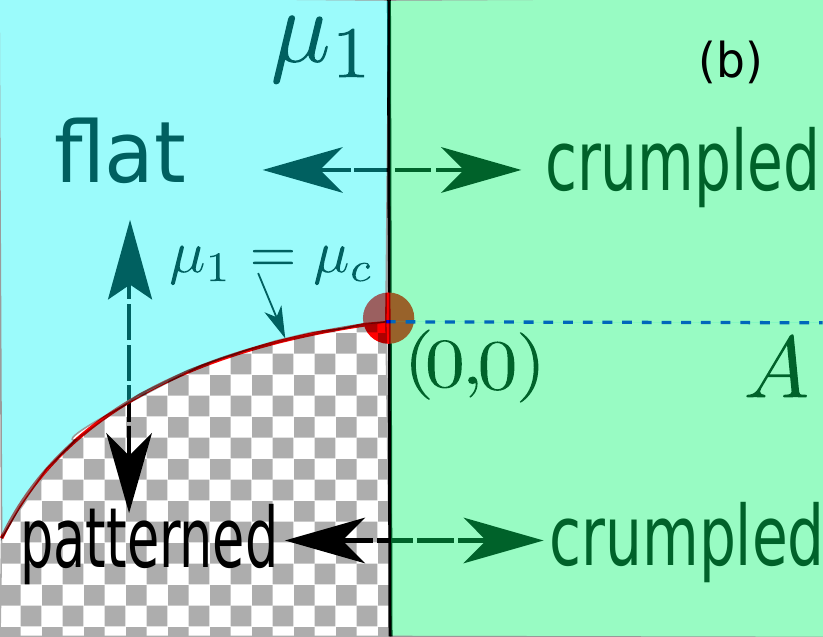}
 %\includegraphics[width=\columnwidth]{phases.pdf}\vskip0.2in
 %\includegraphics[width=\columnwidth]{phase1.pdf}
 %\caption{Phases of active membranes (a)  as  $A$ is varied ($\mu_1>0$): $A<0$ gives the statistically flat phase with orientational LRO and positional QLRO, $A>0$ gives the crumpled phase with orientational and positional short range order and $A=0$ (filled red circle) is the transition point with orientational LRO and positional short range order. (b) as $\mu_1$ is varied ($A<0$): $\mu_1>(<)0$ gives uniform (patterned) states; $\mu_1=0$ (filled red circle) is the threshold of pattern formation. See text.} 
 \caption{(a) Schematic diagram showing active permeation flow  stabilizing ($\mu_1>\mu_c$; green arrows)/destabilizing ($\mu_1<\mu_c$; blue arrows) the membrane. (b)Different phases of symmetric active membranes in the $A-\mu_1$ plane. Three distinct transitions are shown by double arrows, with a triple point at $(0,0)$ (filled red circle).}
 \label{mem-phases}
\end{figure}

%{\em Linear stability:-} 
%We %now study the linear stability of the hydrodynamic equations (\ref{h-eq-lin}), (\ref{uperp-eq-lin}) and (\ref{u-parallel-linear}). Clearly, 
Notice that Eqs.~(\ref{uperp-eq-lin}) and (\ref{u-parallel-linear}) are {\em always} linearly stable, since $\mu>0$ 
and $\lambda+2\mu>0$ for the thermodynamic stability of ${\cal F}$. In contrast, Eq.~(\ref{h-eq-lin}) is linearly 
{ {unstable or stable}} in the long wavelength limit, when the active parameter $A$ is positive (i.e., extensile) 
or negative (i.e., contractile); see Fig.~\ref{model-fig}(c) for a schematic diagram depicting the role of active 
stress in stabilizing ($A<0$)/destabilizing ($A>0$) the membrane: $ {A>0}$ { {enhances the distortion 
of a local patch in the membrane, pushing it locally further from the mean membrane position, whereas $A<0$ 
suppresses the distortions, bringing the patch closer to the  mean position.}  }
Even when { $ {A<0}$}, if $ \mu_1$ is large negative ensuring a sufficient scale separation between { $ {A<0}$} and $ \mu_1<0$ on one hand, and $\mu_1$ and $K$ on the other, $h$-dynamics displays linear instability for an intermediate range of wavevectors. This instability is due to the destabilizing nature of the active permeation flow. { { Qualitatively, the forces on the membranes due to the active proteins which couple with the local curvature, can enhance (suppress) the local curvature, destabilizing (stabilizing) it; see Fig.~\ref{mem-phases}(a) for a pictorial representation.  } } See Fig.~\ref{model-fig}(d) for a schematic plot of $\lambda_h\equiv { -|A|}q/(4\eta') +  |\mu_1| q^2 - Kq^3/(4\eta')$ versus $q$ with $\mu_1<0$, where $\lambda_h$ is the effective damping in (\ref{h-eq-lin}); { $ {A<0}$}. { {The threshold $\mu_c$ of $\mu_1$ for linear instability can be obtained by setting $\lambda_h(q)=0$ and $\partial\lambda_h/\partial q=0$, giving $\mu_c\equiv  - \sqrt{(|A|K)}/{2\eta'}$. Taking $A/(4\eta' a)\sim 1 \,\text{min}^{-1}$~\cite{kruse2006}, $a\sim 100 \text{nm}$ as the typical mesh size in actin cortex~
\cite{cortex-mesh}, we get $A/(4\eta')\sim 100\, \text{nm/min}$; see also Ref.~\cite{basu-noise} for related general discussions. Then using $\eta'\sim 10^{-3}$ kg/(m.sec) and $K\sim 10^{-20}$ J~\cite{chen-perm}, we get a rough estimate $\mu_c\sim 10^{-17}\text{m/sec$^2$}$. On the other hand, dimensionally $\mu_1\sim$ permeation coefficient $\times$ surface tension. Using  the typical values of the parameters~\cite{chen-perm}, we get $\mu_1\sim 10^{-17}\,\text{m/sec$^2$}$, close to $\mu_c$.}} { {Notwithstanding these suggested representative values, in-vitro set ups however allow experimentally controlling these active parameters by a variety of methods, e.g., controlled addition of active proteins and adding or removing of ATP~\cite{girard-biophys,girard-PRL,bouvrais-PNAS,vedia-PNAS,turlier-natphys}}, permitting to explore wider ranges of these parameters.}
%Further, using $\eta'\sim 10^{-3}$ kg/(m.sec) and $K\sim 10^{-20}$ J~\cite{chen-perm}, we get $\mu_c\sim 10^{-17}\text{m/sec$^2$}$, a ballpark figure close to $\mu_1$ close to the instability threshold.}}
%\begin{figure}[htb]
%\includegraphics[width=3.7cm]{active-stress.pdf}\hfill
% \includegraphics[width=3.7cm]{hmode.pdf}
% \caption{ }\label{lambda-h} 
%\end{figure}

%{\em Correlation functions in the linearised hydrodynamics:-} 
To calculate the correlation functions, we now must include noises in the governing dynamical equations. Let $g_h,\,g_\perp$ and $g_\parallel$ be the noises added in Eqs.~(\ref{h-eq-lin}), (\ref{uperp-eq-lin}) and (\ref{u-parallel-linear}), all of which are assumed to be Gaussian-distributed with zero mean. The noises have two sources - thermal and active noises. The thermal noises that survive in the equilibrium limit of the dynamics are controlled by the FDT~\cite{chaikin}, which relates the noise variances to the damping coefficients. This gives zero-mean long range noises with variances~\cite{brochard}
\begin{eqnarray}
 \langle g_h({\bf q},t)g_h({\bf -q},0)\rangle &=& \frac{2k_BT_\text{eff}\delta(t)}{4\eta'q},\label{noise1}\\
 \langle g_\perp({\bf q},t)g_\perp({\bf -q},0)\rangle&=& \frac{2k_BT_\text{eff}\delta(t)}{2\eta'q},\label{noise2}\\
 \langle g_\parallel({\bf q},t)g_\parallel({\bf -q},0)\rangle&=& \frac{2k_BT_\text{eff}\delta(t)}{4\eta'q}, \label{noise3}
\end{eqnarray}
where %$a=\parallel,\,\perp$;  $D_h,\,D_a$  are all positive and scale with $T$
$T_\text{eff}$ is an effective temperature and $k_B$ is the Boltzmann constant. %Unsurprisingly, all the noises are long range due to the long range hydrodynamic damping present even in the equilibrium limit of the dynamics. %The effective temperatures $D_h,\,D_\perp$ and $D_\parallel$ should also scale with the rate of energy released per unit mass by the active processes and the active particle concentrations, making them larger than the ambient thermodynamic temperature. 
Furthermore, there should be additive active noises, which are expected to be short range~\cite{frank-motor,casimir,madan-jacques-PRL,tur2}, 
and hence subdominant to the thermal noises in the long wavelength limit. We thus conclude that although the model is {\em active}, the noises are essentially {\em thermal}. This forms a major conclusion of this work. Do we then have the FDT  in the linearized hydrodynamics? To know this, we must calculate the correlation function of the fields by using (\ref{noise1})-(\ref{noise2}). We consider the linearly stable case with no intermediate wavevector instabilities. In the long wavelength limit, we get
\begin{eqnarray}
 %\langle |h({\bf q},t)|^2\rangle &=&\frac{D_h}{Aq^2},\label{corr-h}\\
 %\langle |u_\perp ({\bf q},t)|^2\rangle &=& \frac{D_\perp}{\mu q^2},\label{corr-uperp}\\
 %\langle |u_\parallel ({\bf q},t)|^2\rangle &=& \frac{D_\parallel}{(\lambda + 2\mu)q^2}.\label{corr-uparallel}
 \langle |h({\bf q},t)|^2\rangle &=&\frac{k_BT}{ {|A|}q^2},\langle |u_a ({\bf q},t)|^2\rangle = \frac{k_BT}{\tilde a q^2}.\label{lin-corr}%,%\langle |u_\parallel ({\bf q},t)|^2\rangle = \frac{D_\parallel}{(\lambda + 2\mu)q^2}.
\end{eqnarray}
Here, $a=\perp,\,\parallel$ and correspondingly $\tilde a={ 2}\mu,\,\lambda+2\mu$, respectively.
Our results in (\ref{lin-corr}) %(\ref{corr-h}), (\ref{corr-uperp}) and (\ref{corr-uparallel}) 
remind us of the known equilibrium results for a fluid membrane with a finite tension  $A$, and a 2D crystal with Lam\'e coefficients $\mu,\lambda$. Nonetheless,  $A$ is an {\em active} dynamical coefficient and not a thermodynamic coefficient, unlike $\mu$ and $\lambda$. The fact that $A$ appears in a static or time-independent quantity $\langle |h({\bf q},t)|^2\rangle$ [see Eq.~(\ref{lin-corr}) above], is a reminder to us that in spite of the apparent equilibrium-like behavior the underlying dynamics is actually a nonequilibrium dynamics. Lastly, Eqs.~(\ref{h-eq-lin}), (\ref{uperp-eq-lin}), (\ref{u-parallel-linear}) and (\ref{lin-corr})
%(\ref{corr-h}), (\ref{corr-uperp}) and (\ref{corr-uparallel}) 
show that all of $h,u_\perp,u_\parallel$ have dynamic exponent $\tilde z=1$ and roughness or spatial scaling exponent $\chi=0$ in 2D, the physically relevant dimension. Here, $\tilde z=1$ means the fluctuations relax much faster than diffusive relaxations. %Due to the active enhancement of $D_h,\,D_a$, fluctuations of $h,u_i$ should be numerically larger in the stable phase than in an equilibrium limit tensed membrane, while displaying the same scaling.

{ {Our theory predicts three distinct phases in this model.
(i)
In the linearly stable case, all of $\langle h^2({\bf r},t)\rangle,\,\langle u_\perp^2({\bf r},t)\rangle,\,\langle u_\parallel ^2({\bf r},t)\rangle$ scale as $\ln(L/a_0)$, giving positional QLRO, where $L$ is the linear system size and $a_0$ is a microscopic cut-off. Likewise, equal-time correlation functions $\langle h({\bf r},t)h(0,t)\rangle,\,\langle u_\perp({\bf r},t)u_\perp (0,t)\rangle,\,\langle u_\parallel ({\bf r},t)u_\parallel (0,t)\rangle$ all behave as $\ln(r/a_o)$, growing logarithmically with $r\equiv |{\bf r}|$, another hallmark of QLRO. Furthermore, $\langle ({\boldsymbol\nabla} h)^2\rangle$ is independent of $L$ for large $L$, giving orientational LRO {{ a statistically flat phase}}. 
(ii)  In the unstable case ($A>0$), independent of $\mu_1$, clearly both $\langle h^2({\bf r},t)\rangle$ and $\langle ({\boldsymbol\nabla} h)^2\rangle$ diverge as soon as $L$ exceeds a threshold determined by the zero of $\lambda_h$ for a finite $q$. This implies crumpled phase of the membrane with both positional and orientational short range order, different from the well-known thermal crumpling phenomenon of equilibrium fluid membranes at any finite $T$. Thermal crumpling occurs in tensionless fluid membranes, whereas in the present case it is due to a negative ``active tension'' $A>0$. What happens if $A=0$? Since $A$ can be varied by varying the strength of the active stress, let us assume it has been tuned to zero, which is the transition point between the ``ordered'' statistically flat phase ($A<0$) and ``disordered'' crumpled phase ($A>0$). At this transition point,  assuming $ \mu_1>0$, we obtain
\begin{equation}
 \langle h^2({\bf r},t)\rangle\sim L,\;\langle ({\boldsymbol\nabla} h)^2\rangle\sim \text{independent of}\,L,
\end{equation}
corresponding to positional short range order and orientational long range order. Thus at the transition point, the membrane is clearly {\em less} ordered than in the ``ordered'' flat phase, but more ordered than the ``disordered'' crumpled phase. %; see Fig.~\ref{mem-phases}(a) for a schematic diagram giving the membrane phases as $A$ is tuned.
Further, $\langle {\bf u}^2({\bf x},t)\rangle\sim \ln (L/a_0)$ at $A=0$, giving in-plane positional QLRO.  
With $A=0$, there is no active tension and the membrane is tensionless, nonetheless with $\mu_1>\mu_c(=0$ now), we define effective wavevector-dependent bend modulus $K_\text{eff}\equiv K+\mu_1/q$, which diverges strongly for small $q$, rendering thermal crumpling ineffective. (iii)  
One can observe yet another transition if $A<0$ (i.e., long wavelength stability) and $\mu_1$, another active parameter, is varied: as $\mu_1$ becomes less than $\mu_c$, patterns are formed with a preferred wavevector $q_c=\sqrt{|A|/K)}$. The phase boundary between the flat and patterned phases is given by $\mu_1=\mu_c\equiv - \sqrt{|A|K}/(2\eta')$. With relevant values of $\mu_1\sim \mu_c$, a typical experimental realization should be close to the threshold of patterned states. See Fig.~\ref{mem-phases}(b) for a phase diagram  in the $A-\mu_1$ plane.}}

{ {Membrane density fluctuations $\delta \rho$, a non-hydrodynamic variable, are controlled by $v_\parallel$, or $u_\parallel$, with $\langle |\delta \rho({\bf q},t)|^2\rangle$ is {\em finite} in the long wavelength limit for $ {A<0},\,\mu_1>0$, but has divergences if $ {A>0}$, or at finite $q$, if $\mu_1<\mu_c$; see SM~\cite{sm}. This may be experimentally explored by particle tracking~\cite{ann-rev}.}}

%{\em Nonlinear effects:-} 
We now discuss the nonlinear effects so far neglected in our studies above. There are two distinct sources of nonlinearities in the present model - those that exist in equilibrium and appear from $\cal F$ and another class which appear in the nonequilibrium dynamics and have their origin in the underlying active processes. %Some of these nonlinear terms have a geometric origin arising from the definition of the strain tensor (\ref{strain}), while some others have purely dynamical (active) origin. 
{ { These nonlinear terms appear both in the dynamics of $h$ and ${\bf u}$; see SM~\cite{sm} for details.  However, 
symmetries of $h$ and $\bf u$ demand that every additional factors of $h$ and $\bf u$ must be associated with one or more additional gradient operators. Then 
scaling and renormalization arguments together with the linear theory scaling exponents $\chi=0,\,\tilde z=1$ ensure that all these nonlinear terms are {\em irrelevant} in the linearly stable states; see SM~\cite{sm}.}} This means the scaling of the correlation functions in the linearly stable case as reported above are in fact {\em exact} in the asymptotic long wavelength limit.  { { Indeed, since all the potential nonlinear terms are {\em irrelevant} and hence vanish in the long wavelength limit, the theory is {\em effectively linear} in that limit and the nonlinearities play no role in sustaining the order. This is unlike most ordered active matter systems~\cite{toner-tu1,toner-tu2,astik-xy1,astik-xy2,debayan1}, where nonlinear effects are essential, with rare exceptions being those studied in Refs.~\cite{bottom1,bottom2}.} }

The nonlinearities are however expected to be important in two other situations: (i) With $A>0$, the linearized $h$-dynamics given by (\ref{h-eq-lin}) is linearly unstable. However, the dynamics of $u_\perp,\,u_\parallel$ remain linearly stable, as both $\mu$ and $\lambda+2\mu$ are assumed to be always positive. Thus the undulation modes get linearly unstable, but the in-plane displacements remain linearly stable, a situation that is physically unacceptable. After all, with the undulation modes getting bigger in the linearly unstable regime, the strain $u_{ij}$ must also get bigger, eventually affecting the in-plane displacements. This apparent contradiction between this physically acceptable expectation and the linear stability of (\ref{uperp-eq-lin}) and (\ref{u-parallel-linear}) even when $A>0$ can be reconciled by the nonlinear effects. As shown in SM~\cite{sm}, these nonlinearities in (\ref{uperp-eq-lin}) and (\ref{u-parallel-linear}) involve $\nabla_i h$, which also gets bigger with $A>0$, which in turn makes fluctuations in ${\bf u}$ growing leading to {\em nonlinear instabilities}. (ii) The exponential growth of $h$ in the unstable case with $A>0$ (or with $\mu_1<\mu_c$) ultimately must saturate due to the nonlinear effects. This is outside the scope of the present work.

%{\em Bulk velocity solutions:-} 
As the membrane fluctuates, it stirs the embedding bulk fluid, generating bulk flows. This can be measured, which calls for calculation of $v'_\alpha$ as a function of $z$. Solving the 3D Stokes equation for $z>0,\,z<0$,
we find~\cite{sm}
 %These solutions allow us to calculate the correlations of $v_\alpha$. For instance
\begin{eqnarray}
 \langle |v'_\perp({\bf q},z,t)|^2\rangle &=& \frac{\mu^2}{4\eta'^2}q^2 \langle |u_\perp({\bf q},t)|^2\rangle \exp (-2qz),\\
 %\langle |v'_\parallel({\bf q},z,t)|^2\rangle &=&\\
 \langle|v'_z({\bf q},z,t)|^2\rangle &=&\bigg[\frac{A^2q^2}{16\eta'^2}(1-qz)^2 \langle |h({\bf q},t)|^2\rangle\nonumber \\&&+ q^2z^2\frac{(\lambda+2\mu)^2}{16\eta'^2}\bigg]\exp(-2qz)
\end{eqnarray}
for $z>0$. Correlation of $v'_\parallel ({\bf q},z,t)$ can be calculated  from those of $v'_z({\bf q},z,t)$. Similar procedure yields $v'_\alpha$ for $z<0$~\cite{sm}. % by using incompressibility of the 3D fluid, and hence its correlation can be calculated. 
Thus the correlations of $v_\alpha({\bf q},z,t)$ are expressed in terms of the correlations of $h({\bf q},t)$ and $\bf u({\bf q},t)$. Since the latter correlations diverge in the unstable case, the bulk flow should also accordingly be stirred strongly. This can be measured by tracking passive tracer particles~\cite{ann-rev}, giving valuable information about the membrane itself.

 We have thus presented a comprehensive hydrodynamic
theory of a nearly flat active tethered membrane immersed in a bulk isotopic active fluid.  This theory makes quantitative, experimentally testable predictions
on the nature of order and scaling of the in-plane and out of plane fluctuations characterized by orientational LRO and positional QLRO, by using the linearized hydrodynamic theory for the linearly stable ordered phase, which are  {\em exact} in the asymptotic long wavelength limit.  { In case of long wavelength instability, the membrane is crumpled with positional and orientational short range order.} { {For long wavelength stability but strong destabilizing permeation flows, our theory speculates pattern formation. It will be interesting to explore its connections with observed finite size structures in membranes~\cite{pattern-mem1}. }}  { { While our theory applies strictly to inversion-symmetric isotropic membranes, it should provide a stepping stone for a more general theory, accounting for asymmetry, anisotropy and spontaneous curvature, relevant for biomembranes.}} The nature of transitions and the role of fluctuations at the various transitions in this model remain theoretically interesting questions. We expect experiments on {\em in-vitro} set ups involving, say, a graphene sheet coated with identical lipid layers on both sides and immersed in an active fluid bath (e.g., a solution of live orientable bacteria or actin filaments) in its isotropic phase, or in reconstituted membrane-tethered actin cortices~\cite{tur1,jitu} can verify broad features of our theory.

%\section{Active symmetric membranes}

%\section{Summary and outlook}

%\section{Acknowledgement}

{\em Acknowledgement:-} S.M. and A.B. are grateful to J.-F. Joanny and P. Bassereau for their helpful suggestions. The authors thank
%the SERB, DST (India) for partial financial support through the TARE scheme [file no.: TAR/2021/000170] and 
the Alexander von Humboldt Stiftung, Germany for partial financial support through the Research Group Linkage Programme (2024). %A.B. thanks 
%the SERB, DST (India) for partial financial support through the MATRICS scheme [file no.: MTR/2020/000406].

\pagebreak[4]

\newpage

\widetext

\begin{center}
 \large \bf Supplemental Material\\
% \large Active effects can stabilise of destabilise symmetric membranes\\
%Flat or crumpled: states of active symmetric membranes
\end{center}

%\appendix

\section{Derivation of the hydrodynamic equations}\label{hydro-deri}

We derive the linear hydrodynamic equations given in the main text for $h,\,u_\parallel$ and $u_\perp$ here. We closely follow Refs.~\cite{kramer-sm,niladri-epje4-sm} in our derivation outlined below.
Assume a nearly flat membrane patch to be spread out along the XY plane and for concreteness $z=0$ to be its average position.
We consider small fluctuations. The three-dimensional (3D) flow field $v_\alpha^\prime$ satisfies the 3D Stokes equation
$\nabla_\beta\sigma^\prime_{\alpha\beta}=0$ giving
\begin{equation}
 \eta'\nabla_3^2 v_\alpha^\prime - \nabla_\alpha P^\prime =0,
\end{equation}
where $\nabla_\beta\sigma_{\alpha\beta}^\prime$ is given in the main text; $\nabla_3^2\equiv \partial^2/\partial x_\beta^2$ is the 3D Laplacian. Similarly, in the Stokesian limit, the 2D flow field $v_i$ satisfies $\nabla_j \sigma_{ij}^\text{tot} + f_i=0$, where $f_i$ is the shear force due to the bulk flow on the 2D flow in the membrane.

We solve $v_\alpha^\prime$ after Fourier transforming in the in-plane coordinates. It is convenient to resolve $v_\alpha^\prime$ into three components 
$ v_z^\prime,\,v_\perp^\prime = (\hat z \times \hat {\bf q})\cdot {\bf v}^\prime$ and $v_\parallel^\prime =\hat {\bf q}\cdot {\bf v}^\prime$, where ${\bf q}$ is an in-plane Fourier wavevector. These three components then satisfy
\begin{eqnarray}
 \eta^\prime (-q^2 + \partial_z^2)v_\perp^\prime ({\bf q},z,t) &=& 0,\\
 \eta^\prime (-q^2 + \partial_z^2)v_\parallel^\prime ({\bf q},z,t) &=& i qP^\prime ({\bf q},z,t),\\
 \eta^\prime (-q^2 + \partial_z^2)v_z^\prime ({\bf q},z,t) &=&\frac{\partial P^\prime}{\partial z} ({\bf q},z,t).
\end{eqnarray}
%Here $\bf q$ is an in-plane Fourier wavevector.
Now impose incompressibility condition on the 3D bulk velocity, which in the real space is ${\boldsymbol\nabla}\cdot {\bf v}^\prime =0$. This in turn gives $\nabla^2 P^\prime=0$ as the equation for pressure $P^\prime$ in the real space.

Subject to the boundary conditions $v_\alpha'({\bf q},z,t)\rightarrow 0$ as $z\rightarrow \pm\infty$, 
we write the general solutions for $v_\alpha^\prime ({\bf q},z,t)$ and $P^\prime ({\bf q},z,t)$ as
\begin{eqnarray}
 v_\perp^\prime ({\bf q},z,t) &=& C_1({\bf q},t) \exp (-q\,z),\;z>0,\label{v-perp+}\\
                              &=&  C_2({\bf q},t) \exp (q\,z),\;z<0,\label{v-perp-}\\
v_z^\prime ({\bf q},z,t) &=& [A_1({\bf q},t) + z\,B_1({\bf q},t)]\exp (-q\,z),\,z>0,\label{vz+} \\
           &=& [A_2({\bf q},t) + z\,B_2({\bf q},t)]\exp (q\,z),\,z<0,\label{vz-}\\
 P^\prime ({\bf q},z,t) &=& D_1({\bf q},t) \exp(-q\,z),\;z>0,\label{P+} \\
          &=& D_2({\bf q},t) \exp(q\,z),\;z<0. \label{P-}
\end{eqnarray}

Lastly, $v_\parallel^\prime$ can be solved by using the 3D incompressibility condition giving
\begin{equation}
 v_\parallel^\prime ({\bf q},z,t)= \frac{i}{q}\frac{\partial v_z^\prime ({\bf q},z,t)}{\partial z}.
\end{equation}
Using the continuity of the 3D bulk flow velocity at $z=h\approx 0$, we get
\begin{equation}
 C_1=C_2,\;A_1=A_2,\; \text{and}\;-qA_1 +B_1 = qA_2 +B_2.
\end{equation}
Then using the $z$-component of the 3D Stokes equation, we find
\begin{equation}
 D_1=2\eta' B_1,\;D_2=2\eta' B_2.
\end{equation}
Next, we find
\begin{equation}
 (2\eta' \frac{\partial v_z'}{\partial z}-P')|_{z=0_+}-(2\eta' \frac{\partial v_z'}{\partial z}-P')|_{z=h_-} =- 4\eta'q A_1 = -4\eta'qv_z(z=h),\label{SM-normal}
\end{equation}
where we have used $n_z=1$ and $n_i=-\nabla_i h$ to the leading order in $h$-fluctuations.
Then, $\nabla_j \sigma_{zj}^a = A\nabla_jn_zn_j = -A \nabla^2 h$ to the lowest order in ${\boldsymbol\nabla}h$.  
Combining everything, the normal stress balance condition can be used to obtain
\begin{equation}
 \frac{\partial h}{\partial t} -v_\text{perm} = v_z(z=h)= \frac{Aq}{4\eta'}h({\bf q},t) - \frac{1}{4\eta'q}\frac{\delta F}{\delta h},
\end{equation}
giving 
\begin{equation}
 \frac{\partial h}{\partial t} = \frac{Aq}{4\eta'}h({\bf q},t) -\mu_1 q^2 ({\bf q},t) - \frac{1}{4\eta'q}\frac{\delta F}{\delta h},
\end{equation}
which is the linearized hydrodynamic equation for the undulation mode to the lowest order in $h$, as already obtained in the main text.

We now derive the hydrodynamic equations for the in-plane displacements $u_i$ to the linear order in ${\boldsymbol\nabla}h$. It is convenient to resolve $\bf u$ in components parallel and perpendicular to $\bf q$ in the XY plane: ${\bf u}\equiv (u_\parallel,\,u_\perp)$. Resolving the 2D Stokes equation parallel and perpendicular to $\hat {\bf q}$, we get to to the linear order in ${\boldsymbol\nabla}h$
\begin{eqnarray}
 &&-({ 2}\eta +\eta_b) q^2\,v_\parallel ({\bf q},t) - i qP - (\lambda+2\mu) q^2 u_\parallel ({\bf q},t) = - f_\parallel ({\bf q},z= h\approx 0,t), \\
 &&-\eta q^2 v_\perp({\bf q},t) - \mu q^2 u_\perp({\bf q},t) = - f_\perp({\bf q},z=h\approx 0,t),
\end{eqnarray}
to the leading order in $h$-fluctuations. In deriving the above equations, we have used~\cite{chaikin-sm}
\begin{equation}
 \sigma_{ij}^\text{el} = \lambda u_{mm}\delta_{ij} + 2\mu u_{ij}
\end{equation}
as the elastic stress. We find
\begin{equation}
 f_\perp({\bf q},z=h\approx 0,t)=\eta^\prime\left[\frac{\partial v_\perp^\prime({\bf q},t)}{\partial z}|_{z=h_+} - \frac{\partial v_\perp^\prime({\bf q},t)}{\partial z}|_{z=h_-}\right]= -\eta'(C_1+C_2)q = - 2\eta' q v^\prime_\perp.\label{SM-shear}
\end{equation}
Combining, we find in the long wavelength limit
\begin{equation}
 \frac{\partial u_\perp ({\bf q},t)}{\partial t} = v_\perp ({\bf q},t) = - \frac{\mu q_\perp}{2\eta'} u_\perp ({\bf q},t)
\end{equation}
to the linear order in (assumed small) fluctuations, as obtained in the main text.

We proceed in the same way to obtain the hydrodynamic equation for $v_\parallel ({\bf q},t)$. We find
\begin{equation}
 f_\parallel ({\bf q},z=h\approx 0,t)=\eta'\left[\frac{\partial v_\parallel^\prime}{\partial z}|_{z=0_+}- \frac{\partial v_\parallel^\prime}{\partial z}|_{z=0_-}\right]=-4\eta'qv_\parallel ({\bf q},z=0,t).
\end{equation}
This gives in the long wavelength limit
\begin{equation}
 4\eta'q v_\parallel ({\bf q},t) + iqP = - (\lambda+2\mu)q^2 u_\parallel ({\bf q},t).\label{v-parallel-pressure}
\end{equation}
Here, the 2D pressure of the  particles $P=\chi(\rho)$, where $\rho$ is the membrane density. If there are birth-death processes, then $\rho$ is non-conserved, but there is a mean density $\rho_0$. Number density fluctuations $\delta \rho$ about $\rho_0$ are {\em fast} variables, i.e., its relaxation time remains finite even in the limit $q\rightarrow 0$, and hence is non-hydrodynamic and is slaved to $v_\parallel$. This can be shown as follows. Density $\rho({\bf x},t)$ follows a mass balance equation. We consider a simple form 
\begin{equation}
 \frac{\partial \rho}{\partial t} + {\nabla_i (v_i\rho)}= \lambda_g \rho - \lambda_d\rho^2,
\end{equation}
corresponding to an average density $\rho_0\equiv \lambda_g/\lambda_d$; $i=x,y$. Here, $\lambda_g>0,\,\lambda_d>0$ are birth and death (due to over crowding) rates. Linearizing about $\rho_0$, to the linear order in fluctuations we find
\begin{equation}
 \frac{\partial\delta\rho}{\partial t}+\rho_0\nabla_i v_i= -\lambda_g\delta\rho.
\end{equation}
In the long time (or low frequency) limit, we thus find in the Fourier space 
\begin{equation}
 \delta \rho({\bf q},t) = -\frac{\rho_0}{\lambda_g} iqv_\parallel({\bf q},t).\label{den-eq}
\end{equation}
Substituting for $\delta\rho$ in \eqref{v-parallel-pressure}, we note that it only contributes to ${\cal O}(q^2)v_\parallel$ and hence is subleading to the ${\cal O}(qv_\parallel')$ term in \eqref{v-parallel-pressure}.
Therefore, in the spirit of hydrodynamics, we drop $\delta\rho$ from our theory in the large time, long wavelength limit and finally get
\begin{equation}
 \frac{\partial u_\parallel ({\bf q},t)}{\partial t}\equiv v_\parallel ({\bf q},t) = - \frac{\lambda+2\mu}{4\eta'}q u_\parallel ({\bf q},t),
\end{equation}
as obtained in the main text. Furthermore, \eqref{den-eq} gives that $\langle |\delta\rho({\bf q},t)|^2\rangle \sim q^2\langle |v_\parallel ({\bf q},t)|^2\rangle$, which has no divergence in the stable (i.e., $A<0,\,\mu_1>0$) case, but diverges when either or $A$ or $\mu_1$ changes sign. 

\section{Solutions of the bulk flow fields}

We use the general solutions of the 3D Stokes equation to express $v'_\alpha({\bf q},z,t)$ in terms of $h({\bf q},t), u_\parallel({\bf q},t)$ and $u_\perp ({\bf q},t)$. We separately solve in the upper ($z>0$) and lower ($z<0$) half planes. We use \eqref{v-perp+} and \eqref{v-perp-}. We find
\begin{eqnarray}
 v'_\perp ({\bf q},z,t)&=&C_1({\bf q},t)\exp(-qz) = v_\perp ({\bf q},t)\exp(-qz)=-\frac{\mu}{2\eta'}qu_\perp ({\bf q},t)\exp(-qz),\;\;z>0\label{vperpprime+}\\
 &=&C_2({\bf q},t)\exp(qz)= v_\perp ({\bf q},t)\exp(qz)=-\frac{\mu}{2\eta'}qu_\perp ({\bf q},t)\exp(qz),\;\;z<0.\label{vperpprime-}
\end{eqnarray}
We now solve for $v_z'({\bf q},z,t)$. We note that
\begin{equation}
 v_z'({\bf q},z=0,t)=A_1({\bf q},t)=A_2({\bf q},t).
\end{equation}
Next, by using the results obtained in Section~\ref{hydro-deri}, we find
\begin{equation}
 B_1-B_2=q(A_1+A_2)=2qv_z'({\bf q},z=0,t),\;\;\;B_1+B_2= - 2iqv'_\parallel ({\bf q},z=0,t)=-2iqv_\parallel({\bf q},t).
\end{equation}
Solving
\begin{eqnarray}
 B_1&=& qv_z'({\bf q},z=0,t) - i qv_\parallel({\bf q},t),\\
 B_2&=& -qv_z'({\bf q},z=0,t) - i qv_\parallel({\bf q},t).
\end{eqnarray}
We thus obtain
\begin{eqnarray}
 v'_z({\bf q},z,t)&=&[v'_z({\bf q},z=0,t)+z(qv'_z({\bf q},z=0,t)-iqv_\parallel({\bf q},t))]\exp(-qz)\nonumber \\&=&\frac{Aq}{4\eta'}h({\bf q},t) (1+qz)\exp(-qz)+iq^2z\frac{\lambda+2\mu}{4\eta'}u_\parallel ({\bf q},t) \exp(-qz),\,\,z>0\label{vzprime+}\\
 v'_z({\bf q},z,t)&=&[v'_z({\bf q},z=0,t)-z(qv'_z({\bf q},z=0,t)-iqv_\parallel({\bf q},t))]\exp(qz)\nonumber \\&=&\frac{Aq}{4\eta'}h({\bf q},t) (1-qz)\exp(-qz)+iq^2z\frac{\lambda+2\mu}{4\eta'}u_\parallel ({\bf q},t) \exp(qz),\,\,z<0.\label{vzprime-}
\end{eqnarray}
to the lowest order in (assumed small) fluctuations.
Finally, $v_\parallel'({\bf q},z,t)=(i/q)\partial v_z'({\bf q},z,t)/\partial z$ due to the incompressibility of the embedding bulk fluid. This gives
\begin{eqnarray}
  v_\parallel^\prime({\bf q},z,t)&=&-iq^2z\frac{A}{4\eta'}h({\bf q},t)\exp(-qz)+(q^2z-q)\frac{\lambda+2\mu}{4\eta'}u_\parallel ({\bf q},t) \exp(-qz),\,\,z>0\label{vparallelprime+}\\
   &=& -iq^2z\frac{A}{4\eta'}h({\bf q},t)\exp(qz)-(q^2z+q)\frac{\lambda+2\mu}{4\eta'}u_\parallel ({\bf q},t) \exp(qz),\,\,z>0.\label{vparallelprime-}
\end{eqnarray}

%We find in the upper half plane with $z>0$
%\begin{eqnarray}
% v^\prime_z(q,z,t)&=&\frac{A}{4\eta'}h({\bf q},t) (1+qz)\exp(-qz)\nonumber \\&+&iqz\frac{\lambda+2\mu}{4\eta'}qu_\parallel ({\bf q},t) \exp(-qz),\label{vzprime}\\
% v_\parallel^\prime(q,z,t)&=&-iqz\frac{A}{4\eta'}h({\bf q},t)\nonumber \\&-&(q-q^2z)\frac{\lambda+2\mu}{4\eta'}u_\parallel ({\bf q},t) \exp(-qz),\label{vparallelprime}\\
% v_\perp^\prime ({\bf q},t) &=& -\frac{\mu}{2\eta'}qu_\perp ({\bf q},t)\exp(-qz).\label{vperpprime}
%\end{eqnarray}
%To get the solutions of $v_\alpha$ in the lower half plane $z<0$, replace $z$ by $-z$ in the above solutions.

\section{Nonlinear terms}

We now systematically derive the nonlinear terms which may be added to the dynamical equations for $h,u_\parallel$ and $u_\perp$. There are several sources of possible nonlinearities. We first consider those which are spatially local. These can come from (i) $v_\text{perm}$ as given in the main text, which has both active and equilibrium contributions,  (ii)  $(\delta {\cal F}/\delta h) $ that contributes to equilibrium relaxation, and (iii) active stress $\sigma_{ij}^a$. To proceed further, we start from the form of the nonlinear stress~\cite{nelson-sm}:
\begin{equation}
 u_{ij} = \frac{1}{2}(\partial_i u_j +\partial_j u_i + \partial_i h\partial_j h),
\end{equation}
 neglecting the the contributions quadratic in $\bf u$~\cite{nelson-sm}. This gives
\begin{eqnarray}
 && u_{mm}^2 = ({\boldsymbol\nabla}\cdot {\bf u})^2 + {\boldsymbol\nabla}\cdot {\bf u}\, ({\boldsymbol\nabla} {h})^2 + \frac{1}{4} ({\boldsymbol\nabla}h)^4,\label{an-F}\\
 && u_{ij}u_{ij}= { {\frac{1}{2}}}(\partial_i u_j)^2 +{ {\frac{1}{2}}}(\partial_iu_j)(\partial_ju_i)+\frac{1}{4} ({\boldsymbol\nabla}h)^4 + \frac{1}{2} (\partial_i u_j + \partial_j u_i)\, \partial_i h \partial_j h, \label{an-F-2}
\end{eqnarray}
where we have ignored any total derivative terms. The terms listed in (\ref{an-F}) and (\ref{an-F-2}) appear in the free energy $\cal F$. We then get
\begin{eqnarray}
&& \frac{\delta {\cal F}}{\delta h}|_{\text{nonlinear}}= -(\mu+\frac{\lambda}{2})\nabla_j [({\boldsymbol\nabla}h)^2\nabla_j h]  - \mu\nabla_j [(\partial_i u_j +\partial_j u_i)\partial_i h] - \lambda\nabla[({\boldsymbol\nabla}\cdot {\bf u})\, {\boldsymbol\nabla}h]. \label{F-nonlin}
\end{eqnarray}
The terms in (\ref{F-nonlin}) contribute to the hydrodynamic equation of $h$ through $v_\text{perm}$  and the relaxational term  $(\delta {\cal F}/\delta h)\,1/(4\eta' q) $; see main text. Now consider the nonlinear contributions from the active stress. The active force from the active stress is $\sim \partial_j\sigma_{\alpha j}^a$ and should have nonlinear contributions. This gives a contribution $~q[{\boldsymbol\nabla} h({\boldsymbol\nabla h})^2]_{\bf q}$ in $\partial h/\partial t$, where a subscript $\bf q$ implies ``expressed in Fourier space''.

Next, we consider the nonlocal nonlinear terms. These will originate from the solutions of the 3D bulk flow fields beyond the linear order in $h$. See, e.g., Eq.~(\ref{SM-normal}), which we have evaluated by setting $z=h \approx 0$ that produces the correct results to the linear order in $h$. Going beyond the linear order results would produce factors of $qh$ and its higher powers appearing multiplicatively with the various terms in the linear theory. 

We now consider the possible nonlinear terms in the $\bf u$-equation. As for the undulation modes, the nonlinearities can be local or nonlocal. We first consider the local nonlinear terms. These originate from anharmonic contributions in the free energy $\cal F$. We find
\begin{equation}
 \frac{\delta {\cal F}}{\delta u_i}|_{\text{nonlinear}} = -\mu \nabla_j [\nabla_i h\,\nabla_j h]- \frac{\lambda}{2} \nabla_i ({\boldsymbol\nabla}h)^2.\label{u-local-non}
\end{equation}
Projecting (\ref{u-local-non}) parallel and perpendicular to $\hat {\bf q}$ in the $XY$ plane produces the local nonlinear contributions to the hydrodynamic equations for $u_\parallel$ and $u_\perp$. There should be nonlinear contributions from the active stress as well. In particular, both $v_\parallel({\bf q},t),\,v_\perp({\bf q},t)$, and hence $\partial u_\parallel/\partial t,\,\partial u_\perp/\partial t$ will have terms $\sim ({\boldsymbol\nabla}h)^2_{\bf q}$, where a subscript $\bf q$ implies ``expressed in Fourier space''. Finally, similar to the undulation modes, there are nonlocal nonlinear contributions to the equations for $u_\parallel$ and $u_\perp$, which originate from the solutions of the 3D ambient flow fields. See, e.g., Eq.~(\ref{SM-shear}) above. Going beyond the linear order in fluctuations produce factors of $qh$ and its higher powers appearing multiplicatively with the terms in the linear hydrodynamic equations for $u_\parallel$ and $u_\perp$. We now find out whether or not these nonlinear terms can change the predictions from the linear hydrodynamic equations in the long wavelength limit. 

To asses the renormalization group (RG) relevance of these above nonlinear terms, we rescale space, time and the fields as
\begin{eqnarray}
 {\bf r'}=b{\bf r},\,t'=b^z t,\,h({\bf r},t)=b^{\chi_h}h({\bf r'},t'),\,u_\perp({\bf r},t)=b^{\chi_\perp}u_\perp ({\bf r'},t'),\,u_\parallel({\bf r},t)=b^{\chi_\parallel}u_\parallel ({\bf r},t).\label{rescale}
\end{eqnarray}
In the linear theory, $z=1,\,\chi_h=\chi_\perp=\chi_\parallel=0$ in 2D, such that with these choices for the scaling exponents, the parameters in the linearised hydrodynamic equations for $h,u_\perp$ and $u_\parallel$ and the corresponding noise strengths remain unchanged under rescaling. Notice that in {\em each} possible nonlinear term as listed above, every factor of a field, $h,\,u_\parallel,\,u_\perp$, is necessarily multiplied by a factor of wavevector. While this can be seen explicitly from the explicit structure of the nonlinear terms, it is expected to be so to ensure invariance under constant shifts of $h, u_i$.  Simple power counting however shows that under such a rescaling procedure, all the nonlinear terms acquire scale factor of $b$ raised to  negative powers: Every additional factor of a field accompanied with a spatial gradient then contributes $b^{\chi_a}b^{-1}\sim b^{-1},\,a=h,u_\parallel,u_\perp$ as a scale factor in 2D under rescaling (\ref{rescale}), where we have used $\chi_a=0$ in 2D. This negative scaling dimension naturally renders each of them irrelevant in the long wavelength limit, at least if they are initially small, making our linear theory {\em exact} in the asymptotic long wavevlength limit.  
%\end{widetext}

\end{document}